\documentclass{article}

\usepackage{arxiv}

\usepackage[utf8]{inputenc}
\usepackage[T1]{fontenc}
\usepackage{hyperref}
\usepackage{url}
\usepackage{booktabs}
\usepackage{amsfonts}
\usepackage{nicefrac}
\usepackage{microtype}
\usepackage{graphicx}

\title{Augmenting Polish Automatic Speech Recognition System With Synthetic Data}

\author{
  \L{}ukasz Bondaruk\\
  Samsung R\&D Institute Poland\\
  Plac Europejski 1\\
  00-844 Warsaw, Poland\\
  \texttt{l.bondaruk@samsung.com}\\
  \And
  Jakub Kubiak\\
  Samsung R\&D Institute Poland\\
  Plac Europejski 1\\
  00-844 Warsaw, Poland\\
  \texttt{j.kubiak@samsung.com}\\
  \AND
  Mateusz Czy\.{z}nikiewicz\\
  Samsung AI Center Warsaw\\
  Plac Europejski 1\\
  00-844 Warsaw, Poland\\
  \texttt{m.czyznikiew@samsung.com}\\
}

\begin{document}
\maketitle

\begin{abstract}
  This paper presents a system developed for submission to Poleval
  2024, Task 3: Polish Automatic Speech Recognition
  Challenge\footnote{\url{http://poleval.pl/tasks/task3}\label{task3}}.
  We describe Voicebox-based speech synthesis pipeline and utilize it
  to augment Conformer and Whisper speech recognition models with synthetic
  data. We show that addition of synthetic speech to training improves
  achieved results significantly. We also present final results achieved by
  our models in the competition.
\end{abstract}


\section{Introduction}
Automatic Speech Recognition (ASR) systems became essential in enabling machines to understand
and process human language. However, training these systems, especially for less widely
spoken languages like Polish, presents challenges due to the limited availability of
high-quality labeled data. To address this issue, and based on recent advencements
in speech synthesis systems, we decided to explore the use of synthetic data to augment
real datasets. This paper focuses on enhancing Polish ASR by incorporating
synthetic speech into training data, aiming to overcome the scarcity of natural
speech resources and boost model performance. We test our solution on Poleval
2024, Task 3: Polish Automatic Speech Recognition\textsuperscript{\ref{task3}}.

\section{Related Work}
Speech recognition for Polish language has been the subject of many scientific articles. Researchers
have explored classical approaches, such as HMM~\cite{asr-pl-htk}, developed custom systems based on kNN
classifier and wavelets~\cite{asr-pl-knn} and, based on Kaldi toolkit~\cite{kaldi}, trained more
robust systems~\cite{asr-pl-sarmata2, asr-pl-clarin}. Topics of adapting models to specific domain~\cite{asr-pl-games}
or to specific acoustic conditions~\cite{asr-pl-polska-kronika-filmowa} have also been explored.

In parallel with research on speech recognition systems, an increasing number of speech corpora
in Polish have been made available~\cite{datasets-pl-corpora, datasets-pl-jurisdic, datasets-pl-luna}.
Some of the most notable ones include AGH Corpus of Polish Speech~\cite{datasets-pl-agh}
(more than $25$ hours of data from $166$ speakers) and CLARIN-PL~\cite{asr-pl-clarin}
(around $56$ hours of data from $317$ speakers).

With the shift to a multilingual paradigm enabled by the use of deep learning, Polish
began to be included in large multilingual speech corpora such as Multilingual LibriSpeech~\cite{mls} or
VoxPopuli~\cite{voxpopuli}. This led to the development of speech recognition systems where Polish
was just one of the available languages. Some of the most notable ones are
Wav2Vec 2.0 XLSR-53~\cite{wav2vec2-xlsr} and Whisper~\cite{whisper}. More comprehensive
survey of speech corpora for Polish can be found in~\cite{datasets-pl-bigos}.

Topic of speech recognition for Polish has also been tackled in Task 5: Automatic Speech
Recognition of PolEval 2019. The goal was to build a system for transcribing sessions of the
Polish Parliament. Systems developed by four participating teams achieved results ranging
from $41.8\%$ to $11.8\%$ WER.~\cite{poleval19}

Prior work demonstrated that using synthetic data can significantly improve
speech recognition performance. It was shown that augmentation
with synthetic speech can increase the robustness of ASR training, leading
to a $38\%$ relative improvement in some systems~\cite{rossenbach21comparing}. In low-resource
languages, adding synthetic data reduced WER by up to $25.5\%$~\cite{bartelds-etal-2023-making}.

However, augmentation with synthetic speech presented challenges due to the
differences between synthetic and real data. This problem was addressed with the
development of zero-shot voice-cloning TTS systems such as VALL-E~X~\cite{vallex}
and Voicebox~\cite{voicebox}. Authors of Voicebox compared performance of ASR
systems trained on real and synthetic data and observed only small reduction in
quality~\cite{voicebox}. Topic of augmentation to a specific domain with synthetic
speech produced with VALL-E X was also researched~\cite{spoken-lang-corp-aug}.

\section{Data}

Data provided by organizers was distributed through Hugging Face and was comprised of two parts:
BIGOS dataset\footnote{\url{https://huggingface.co/datasets/amu-cai/pl-asr-bigos-v2}}~\cite{datasets-pl-bigos}
and PELCRA benchmark\footnote{\url{https://huggingface.co/datasets/pelcra/pl-asr-pelcra-for-bigos}}.
BIGOS is a compilation of 12 open datasets whereas PELCRA was built as a compilation of selected datasets
from PELCRA repository~\cite{pelcra-tools}. Details on sizes of specific dataset splits are available
in Table~\ref{tab:data}. More detailed summary, with information on specific subsets (data sources)
was provided by the organizers for both parts and all splits, we do not include this information here.

\begin{table}[ht]
  \caption{Summary of dataset split sizes.}
  \label{tab:data}
  \centering
  \begin{tabular}{lrrrrrr}
    \toprule
    & \multicolumn{3}{c}{\textbf{Number of samples}} & \multicolumn{3}{c}{\textbf{Duration [h]}} \\
    \cmidrule(r){2-4} \cmidrule(r){5-7}
    \textbf{Split} & \textbf{BIGOS} & \textbf{PELCRA} & \textbf{Total} & \textbf{BIGOS} & \textbf{PELCRA} & \textbf{Total} \\
    \midrule
    \textit{train} & $82025$ & $229150$ & $311175$ & $236.70$ & $432.26$ & $668.96$ \\
    \textit{dev-0} & $14254$ & $28532$ & $42786$ & $27.51$ & $49.60$ & $77.11$ \\
    \textit{test-A} & $1002$ & $1167$ & $2169$ & $2.53$ & $2.14$ & $4.67$ \\
    \textit{test-B} & $991$ & $1178$ & $2169$ & $2.48$ & $2.15$ & $4.63$ \\
    \bottomrule
  \end{tabular}
\end{table}

Utilization of multiple datasets ensures high variability of data. In particular, PELCRA provides
spontaneous and conversational speech whereas BIGOS contains audiobook data \cite{mls, mailabs},
read speech recorded with many devices and in multiple acoustic conditions \cite{asr-pl-clarin, commonvoice}
and spontaneous speech. Such diversity poses a difficult challenge for ASR systems but also
ensures comprehensive evaluation of system robustness.

\section{Method}
\subsection{Speech recognition}
As our approach focuses mainly on utilization of synthetic data for augmentation of ASR system, we
decided to use two standard speech recognition models without any modifications. Both models
utilize BPE text tokenization scheme.

Conformer~\cite{conformer} combines the strengths of transformer and convolutional neural
network to capture both global and local dependencies in audio sequences. It does so by modifying
transformer block to include additional convolutional module between standard multi-head attention
and feed-forward layer. By integrating these two architectures, the Conformer
demonstrates competitive performance even with compact models and reaches state-of-the-art
accuracy in speech recognition. We utilize a pre-existing RNN-T
implementation\footnote{\url{https://pytorch.org/audio/main/generated/torchaudio.prototype.models.conformer_rnnt_model.html}}
with RNNTLoss\footnote{\url{https://pytorch.org/audio/stable/generated/torchaudio.transforms.RNNTLoss.html}}.
Our model has $60$M parameteres and is trained from scratch for $500$k steps with effective
batch size of $512$ audio samples. For decoding we utilize beam search with beam size of $10$.

Whisper~\cite{whisper} is a standard encoder-decoder transformer model~\cite{transformer} with
small modifications required for handling audio input. Advantage of using Whisper comes from
large-scale supervised pretraining in multitask setting. In this work we perform full fine-tuning
of \textit{whisper-large-v3}\footnote{\url{https://huggingface.co/openai/whisper-large-v3}} model
which has $1550$M parameters. We run the fine-tuning and then choose checkpoint with the best validation
loss. Fine-tuning utilized effective batch size of $64$ audio samples. For decoding we utilize beam
search with beam size of $4$.

\subsection{Speech synthesis}
We adopted a Voicebox-based~\cite{voicebox} strategy for speech synthesis. It offers state-of-the-art
voice cloning in the resulting speech samples. Moreover, this approach can effectively utilize
audios of suboptimal quality that earlier text-to-speech (TTS) systems could not accommodate. By harnessing these
benefits, we aim to produce a synthetic dataset that closely mirrors the original in both quality
and variability.

To achieve this, we trained a collection of models that work as one system. Models were trained from
scratch using only the data provided by the competition organizers.

Voicebox~\cite{voicebox} is a zero-shot TTS that that leverages flow matching~\cite{flowmatching}. 
It enables the generation of audio conditioned on specific text and prompt audio. During the denoising process, 
Voicebox transforms a Gaussian distribution into the target distribution by solving ordinary
differential equation (ODE) in a fixed number of steps to produce a mel spectrogram. Its architecture is
built on a transformer encoder, enhanced with U-NET-like connections~\cite{u-net} and rotary
embeddings~\cite{rotary}. Additionally, the model underwent an extra pretraining stage in the manner
described in~\cite{audiobox}, this step utilized only audio data. Model was pretrained for $270$k
steps and then adapted for $200$k steps. It has $430$M parameters and during training the
effective batch size of $256$ audio samples was used. For inference, we used $15$ steps
with midpoint ODE solver.

CTCAligner is a module that aligns audio features with text tokens in a force-aligner manner.
This alignment provides information on the duration of each token, allowing for speaker intonation cloning.
It shares the same architecture as Voicebox and utilizes CTC loss, enabling it to generate the mapping
in an unsupervised manner. It also went through pretraining step in Best-RQ manner~\cite{best-rq}. During
this pretraining step audio features are aligned to codes from random frozen codebook. The model
has $36$M parameters. It was pretrained for $1$M steps and then adapted for another
$1$M steps. Effective batch size of $512$ audio samples was utilized.

DurationPredictor~\cite{voicebox} is built in the same way as Voicebox and also takes advantage of flow matching.
Its primary function is to estimate the duration of each target token based on the context provided
by the results of the CTCAligner. This allows to effectively transfer the intonation from the
prompt speech to the target speech. The trained model has $93$M parameters and was trained
for $50$k steps with effective batch size of $8192$. For inference, we used $10$ steps with midpoint
ODE solver, also we calculate average of $10$ model runs.

HiFi-GAN~\cite{hifigan} is a fully convolutional generative adversarial network that functions
as a vocoder, converting mel spectrogram features into audio signals. It employs both multi-scale and
multi-period discriminators, enabling it to achieve exceptionally high-quality audio output. The
model has $14$M parameters and the training was run for $1$M steps with effective batch
size of $512$.

\subsection{Data preparation}

For each model in the speech synthesis pipeline, we utilized the entire \textit{train} split from
the data provided by the organizers. The audio files were processed by extracting mel spectrograms
using the following parameters: sample rate of $16$kHz, hop size of $256$, window length of $1024$,
minimum frequency of $0$kHz, maximum frequency of $8$kHz, and with $80$ mel channels. Text data
was lowercased.

For speech recognition models training, in addition to using the entire \textit{train} split of
data provided by the organizers, we incorporated two synthetic datasets. These were generated
using speech synthesis system applied to randomly selected prompts taken from \textit{train} split that
were filtered based on the output of our \textit{conformer-baseline} recognizer and speech rate
criteria. Only audio files that achieved a maximum character error rate (CER) of $25\%$ and
had a speech rate variation within the range of $0$ to $2.5$ standard deviations from mean
were selected for synthesis. This process resulted in creating two synthetic datasets:
\textit{synth-00} ($440$ hours and $293496$ audio samples) and \textit{synth-01}
($890$ hours and $586992$ audio samples). By mixing these datasets with real data we created
three training datasets, details on their sizes and composition are provided in
Table~\ref{tab:synth_data}.

\begin{table}[ht]
  \caption{Summary of datasets used for speech recognition models' training.}
  \label{tab:synth_data}
  \centering
  \begin{tabular}{llrr}
    \toprule
    \textbf{Dataset} & \textbf{Composition} & \textbf{Number of samples} & \textbf{Duration [h]} \\
    \midrule
    \textit{baseline} & \textit{train} & $311175$ & $669$ \\
    \textit{mix-00} & \textit{train} + \textit{synth-00} & $604671$ & $1109$ \\
    \textit{mix-01} & \textit{train} + \textit{synth-00} + \textit{synth-01} & $1191663$ & $1999$ \\
    \bottomrule
  \end{tabular}
\end{table}

As a form of additional augmentation, for speech recognizers, we applied time and frequency masking
from SpecAugment~\cite{specaug}. Without time warping, which was also proposed in \cite{specaug},
we were able to precompute all mel spectrograms what made training faster. Moreover authors of
SpecAugment suggest that time warping has little to no effect on final results. Based on the
findings in~\cite{comparison-asr-augmentation}, other augmentation techniques, such as adding
noise or speech perturbations, were determined to be of questionable benefit, and therefore,
were not utilized in our study.

\section{Results}

Word error rate (WER) and character error rate (CER) were used for comparing submissions.
WER is defined as the number of incorrectly transcribed words divided by the total number of
words in the reference sentences whereas CER is defined as the number of incorrectly transcribed
characters divided by the total number of characters in the reference sentences. For development
purposes we only utilized WER metric, calculated on \textit{dev-0} split, for these calculations
we lowercased all text and removed all punctuation. All rates are multiplied by $100$ for better
readability.

\begin{table}[ht]
  \caption{Mean word error rate (WER) for all evaluated models calculated on \textit{dev-0} splits for both data sources. Mean is weighted based on number of samples in subsets.}
  \label{tab:results-dev}
  \centering
  \begin{tabular}{lrrr}
    \toprule
    \textbf{Model} & \textbf{BIGOS} & \textbf{PELCRA} & \textbf{Total} \\
    \midrule
    \textit{whisper-large-v3} & $6.08$ & $29.04$ & $21.39$ \\
    \midrule
    \textit{whisper-large-v3-baseline} & $6.16$ & $23.35$ & $17.62$ \\
    \textit{whisper-large-v3-mix-00} & $5.04$ & $22.58$ & $16.74$ \\
    \textit{whisper-large-v3-mix-01} & $3.93$ & $20.98$ & $15.30$ \\
    \midrule
    \textit{conformer-baseline} & $11.22$ & $30.55$ & $24.11$ \\
    \textit{conformer-mix-00} & $7.85$ & $27.32$ & $20.84$ \\
    \textit{conformer-mix-01} & $7.26$ & $25.38$ & $19.34$ \\
    \bottomrule
  \end{tabular}
\end{table}

Results of our internal evaluations are shown in Table~\ref{tab:results-dev}. We provide results
for both Conformer and Whisper models trained on all studied training datasets. For comparison,
we also evaluated Whisper without any fine-tuning. Results confirm that the addition of synthetic
data improves quality of both models. The results are also not clearly saturated even in the case
of \textit{mix-01} where total duration of training data was almost tripled. Addition of synthetic
data seems to have more impact on the results of Conformer model - between \textit{mix-01} and
\textit{baseline}, total WER was reduced by $4.77$, whereas in the case of Whisper it was reduced only
by $2.32$. This can be explained by large-scale pretraining that Whisper did undergo - in its case
the \textit{baseline} achieved significantly better results than Conformer. We can also observe
that in the case of Whisper, fine-tuning has more impact on the results achieved on PELCRA part of data
- between raw model and \textit{mix-01}, WER was reduced by $8.06$ on PELCRA and only by $2.15$ on BIGOS.
This is probably connected to the overall worse results achieved by the models on PELCRA part of
the data, which may be caused by PELCRA being more difficult (conversational and spontaneous speech).
The best results on \textit{dev-0} split were achieved by \textit{whisper-large-v3-mix-01} what made
it the main candidate for our submission.

\begin{table}[ht]
  \caption{Character error rate (CER) and word error rate (WER) for all evaluated models calculated on both \textit{test-A} and \textit{test-B} splits.}
  \label{tab:results-test}
  \centering
  \begin{tabular}{lrrrr}
    \toprule
    & \multicolumn{2}{c}{\textbf{\textit{test-A}}} & \multicolumn{2}{c}{\textbf{\textit{test-B}}} \\
    \cmidrule(r){2-3} \cmidrule(r){4-5}
    \textbf{Model} & \textbf{CER} & \textbf{WER} & \textbf{CER} & \textbf{WER} \\
    \midrule
    \textit{whisper-large-v3-baseline} & $7.15$ & $11.52$ & $7.10$ & $11.23$ \\
    \textit{whisper-large-v3-mix-00} & $6.85$ & $11.07$ & $6.91$ & $11.15$ \\
    \textit{whisper-large-v3-mix-01} & $6.90$ & $11.27$ & $6.85$ & $11.07$ \\
    \midrule
    \textit{conformer-baseline} & $8.77$ & $17.48$ & $8.37$ & $16.82$ \\
    \textit{conformer-mix-00} & $7.60$ & $15.25$ & $7.16$ & $14.33$ \\
    \textit{conformer-mix-01} & $7.08$ & $13.99$ & $6.90$ & $13.40$ \\
    \bottomrule
  \end{tabular}
\end{table}

Results achieved by all our models on \textit{test-A} and \textit{test-B} splits are shown in
Table~\ref{tab:results-test}. As one would expect, all measured character error rates are lower
than corresponding word error rates. Also, results show that addition of synthetic data has smaller
impact on Whisper than in the case of \textit{dev-0} split. This may be caused by test splits being
balanced with regard to data source being BIGOS or PELCRA and the main gains in results were
achieved on PELCRA data. We can also observe that \textit{conformer-mix-01} and
\textit{whisper-large-v3-mix-01} have similar quality when measured with CER (difference of $0.18$
on \textit{test-A} and $0.05$ on \textit{test-B}) but differences in WER are larger ($2.72$ on
\textit{test-A} and $2.33$ on \textit{test-B}). It is also not obvious whether
\textit{whisper-large-v3-mix-00} or \textit{whisper-large-v3-mix-01} is better as one achieves better
results on \textit{test-A} and second one on \textit{test-B}.

\section{Discussion}
It is possible to present legitimate doubts regarding whether the proposed system, that includes
augmentation with synthetic speech, complies with the competition rules. Specifically, with the
provision stating: "It is forbidden for the participants to use any data outside of the provided
train and validation sets to develop their systems". However, we argue that this system meets the
competition requirements, and our reasoning is provided below.

\begin{figure*}[ht]
  \begin{center}
    \includegraphics[scale=0.28]{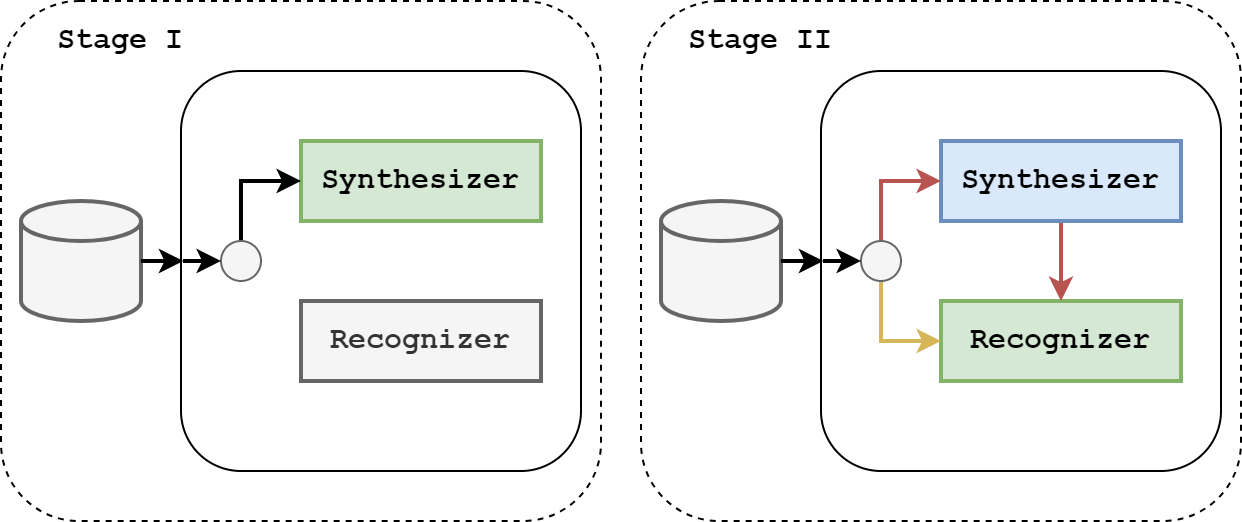}
    \caption{Automatic speech recognition system augmented with synthetic speech presented as a hierarchical system. In~Stage I, Synthesizer is trained, its weights are then frozen in Stage II where Recognizer is trained. In Stage II, data is sampled and provided either directly to the Recognizer (yellow data flow) or first is processed with Synthesizer and only then is provided to Recognizer (red data flow).}
    \label{fig:hierarchical}
  \end{center}
\end{figure*}

First, the entire system, consisting of multiple models, can be viewed as a single
hierarchical system trained within a multi-stage and multi-task paradigm. This approach is illustrated
in Figure~\ref{fig:hierarchical}. Such a system would be trained piece by piece, starting with
synthesizer. Then specific weights would be frozen, and further training procedure would utilize
sampling that would select appropriate path through the system to accurately reproduce the
augmentation process. This is represented in Figure~\ref{fig:hierarchical} as red and yellow
data flow in Stage II. Additionally, certain components of this system could be viewed as
interpretable intermediate points (e.g. output of speech synthesis).

Second, in order to both train and to infer from synthesizer we utilize only data provided by
the organizers of the competition. No external data was used for that purpose. The gained variability
of synthetic speech comes from mixing texts and voices from different samples but we argue that this
does comply with the competition rules.

\section{Conclusions}
We have presented an automatic speech recognition system for Polish augmented with synthetic data
generated using state-of-the-art speech synthesizer. We evaluated two speech recognition models,
both with and without augmentation. One of the models was trained from scratch and second was only
fine-tuned. We presented results we obtained using this approach in Poleval 2024, Task 3: Polish
Automatic Speech Recognition. We showed that introducing augmentation with synthetic speech improves
the system's results.

There are still some avenues for improvements of the proposed system. In order to introduce
more variability in the synthetic data we could utilize language model to generate texts for synthesizer.
More careful procedure for choosing audio prompts for voice cloning could also be introduced. In particular,
by introducing a more iterative model training procedure, we could select prompts from subsets on which
model performs worse. But due to competition time constraints and infrastructure limitations we leave these topics
for further research.

\bibliographystyle{unsrt}
\bibliography{main}

\begin{thebibliography}{10}

\bibitem{asr-pl-htk}
Bartosz Zi{\'o}lko, Suresh Manandhar, Richard~C. Wilson, Mariusz Zi{\'o}lko,
  and Jakub Galka.
\newblock Application of htk to the polish language.
\newblock In {\em 2008 International Conference on Audio, Language and Image
  Processing}, pages 1759--1764, 2008.

\bibitem{asr-pl-knn}
Mariusz Ziółko, Jakub Gałka, Bartosz Ziółko, Tomasz Jadczyk, Dawid
  Skurzok, and Mariusz Masior.
\newblock Automatic speech recognition system dedicated for polish.
\newblock In {\em Proc. Interspeech 2011}, pages 3315--3316, 2011.

\bibitem{kaldi}
Daniel Povey, Arnab Ghoshal, Gilles Boulianne, Lukas Burget, Ondrej Glembek,
  Nagendra Goel, Mirko Hannemann, Petr Motlicek, Yanmin Qian, Petr Schwarz, Jan
  Silovsky, Georg Stemmer, and Karel Vesely.
\newblock The kaldi speech recognition toolkit.
\newblock In {\em IEEE 2011 Workshop on Automatic Speech Recognition and
  Understanding}. IEEE Signal Processing Society, December 2011.
\newblock IEEE Catalog No.: CFP11SRW-USB.

\bibitem{asr-pl-sarmata2}
Bartosz Zi{\'o}łko, Tomasz Jadczyk, Dawid Skurzok, Piotr Żelasko, Jakub
  Gałka, Tomasz Pedzimaz, Ireneusz Gawlik, and Szymon Palka.
\newblock Sarmata 2.0 automatic polish language speech recognition system.
\newblock In {\em Proc. Interspeech 2015}, pages 1062--1063, 2015.

\bibitem{asr-pl-clarin}
Danijel Korzinek, Krzysztof Marasek, Lukasz Brocki, and Krzysztof Wolk.
\newblock Polish read speech corpus for speech tools and services.
\newblock In Lars Borin, editor, {\em Selected papers from the {CLARIN} Annual
  Conference 2016, Aix-en-Provence, France, October 26-28, 2016}, volume 136 of
  {\em Link{\"{o}}ping Electronic Conference Proceedings}, pages 54--62.
  Link{\"{o}}ping University Electronic Press, 2016.

\bibitem{asr-pl-games}
Artur Janicki and Dariusz Wawer.
\newblock Automatic speech recognition for polish in a computer game interface.
\newblock In {\em 2011 Federated Conference on Computer Science and Information
  Systems (FedCSIS)}, pages 711--716, 2011.

\bibitem{asr-pl-polska-kronika-filmowa}
Danijel Koržinek, Krzysztof Wołk, Łukasz Brocki, and Krzysztof Marasek.
\newblock Automatic transcription of the polish newsreel.
\newblock {\em Poznan Studies in Contemporary Linguistics}, 55(2):183--209,
  2019.

\bibitem{datasets-pl-corpora}
Stefan Grocholewski.
\newblock Corpora - speech database for polish diphones.
\newblock In {\em Proc. 5th European Conference on Speech Communication and
  Technology (Eurospeech 1997)}, pages 1735--1738, 1997.

\bibitem{datasets-pl-jurisdic}
Grazyna Demenko, Stefan Grocholewski, Katarzyna Klessa, Jerzy Og{\'o}rkiewicz,
  Agnieszka Wagner, Marek Lange, Daniel {\'S}ledzi{\'n}ski, and Natalia Cylwik.
\newblock {JURISDIC}: {P}olish speech database for taking dictation of legal
  texts.
\newblock In Nicoletta Calzolari, Khalid Choukri, Bente Maegaard, Joseph
  Mariani, Jan Odijk, Stelios Piperidis, and Daniel Tapias, editors, {\em
  Proceedings of the Sixth International Conference on Language Resources and
  Evaluation ({LREC}'08)}, Marrakech, Morocco, May 2008. European Language
  Resources Association (ELRA).

\bibitem{datasets-pl-luna}
Małgorzata Marciniak.
\newblock {\em Anotowany korpus dialogów telefonicznych}.
\newblock EXIT, 2010.

\bibitem{datasets-pl-agh}
Piotr Żelasko, Bartosz Ziółko, Tomasz Jadczyk, and Dawid Skurzok.
\newblock Agh corpus of polish speech.
\newblock {\em Language Resources and Evaluation}, 50(3):585--601, 2016.

\bibitem{mls}
Vineel Pratap, Qiantong Xu, Anuroop Sriram, Gabriel Synnaeve, and Ronan
  Collobert.
\newblock Mls: A large-scale multilingual dataset for speech research.
\newblock In {\em Proc. Interspeech 2020}, pages 2757--2761, 2020.

\bibitem{voxpopuli}
Changhan Wang, Morgane Riviere, Ann Lee, Anne Wu, Chaitanya Talnikar, Daniel
  Haziza, Mary Williamson, Juan Pino, and Emmanuel Dupoux.
\newblock {V}ox{P}opuli: A large-scale multilingual speech corpus for
  representation learning, semi-supervised learning and interpretation.
\newblock In Chengqing Zong, Fei Xia, Wenjie Li, and Roberto Navigli, editors,
  {\em Proceedings of the 59th Annual Meeting of the Association for
  Computational Linguistics and the 11th International Joint Conference on
  Natural Language Processing (Volume 1: Long Papers)}, pages 993--1003,
  Online, August 2021. Association for Computational Linguistics.

\bibitem{wav2vec2-xlsr}
Alexis Conneau, Alexei Baevski, Ronan Collobert, Abdelrahman Mohamed, and
  Michael Auli.
\newblock Unsupervised cross-lingual representation learning for speech
  recognition.
\newblock In {\em Proc. Interspeech 2021}, pages 2426--2430, 2021.

\bibitem{whisper}
Alec Radford, Jong~Wook Kim, Tao Xu, Greg Brockman, Christine McLeavey, and
  Ilya Sutskever.
\newblock Robust speech recognition via large-scale weak supervision, 2022.

\bibitem{datasets-pl-bigos}
Michal Junczyk.
\newblock Bigos - benchmark intended grouping of open speech corpora for polish
  automatic speech recognition.
\newblock In {\em 2023 18th Conference on Computer Science and Intelligence
  Systems (FedCSIS)}, pages 585--590, 2023.

\bibitem{poleval19}
Proceedings of the poleval 2019 workshop.
\newblock Warsaw, Poland, 2019. Institute of Computer Science, Polish Academy
  of Sciences.

\bibitem{rossenbach21comparing}
Nick Rossenbach, Mohammad Zeineldeen, Benedikt Hilmes, Ralf Schlüter, and
  Hermann Ney.
\newblock Comparing the benefit of synthetic training data for various
  automatic speech recognition architectures.
\newblock In {\em IEEE Automatic Speech Recognition and Understanding
  Workshop}, pages 788--795, 2021.

\bibitem{bartelds-etal-2023-making}
Martijn Bartelds, Nay San, Bradley McDonnell, Dan Jurafsky, and Martijn
  Wieling.
\newblock Making more of little data: Improving low-resource automatic speech
  recognition using data augmentation.
\newblock In Anna Rogers, Jordan Boyd-Graber, and Naoaki Okazaki, editors, {\em
  Proceedings of the 61st Annual Meeting of the Association for Computational
  Linguistics (Volume 1: Long Papers)}, pages 715--729, Toronto, Canada, July
  2023. Association for Computational Linguistics.

\bibitem{vallex}
Ziqiang Zhang, Long Zhou, Chengyi Wang, Sanyuan Chen, Yu~Wu, Shujie Liu, Zhuo
  Chen, Yanqing Liu, Huaming Wang, Jinyu Li, Lei He, Sheng Zhao, and Furu Wei.
\newblock Speak foreign languages with your own voice: Cross-lingual neural
  codec language modeling, 2023.

\bibitem{voicebox}
Matthew Le, Apoorv Vyas, Bowen Shi, Brian Karrer, Leda Sari, Rashel Moritz,
  Mary Williamson, Vimal Manohar, Yossi Adi, Jay Mahadeokar, and Wei-Ning Hsu.
\newblock Voicebox: Text-guided multilingual universal speech generation at
  scale.
\newblock In A.~Oh, T.~Naumann, A.~Globerson, K.~Saenko, M.~Hardt, and
  S.~Levine, editors, {\em Advances in Neural Information Processing Systems},
  volume~36, pages 14005--14034. Curran Associates, Inc., 2023.

\bibitem{spoken-lang-corp-aug}
Mateusz Czyżnikiewicz, Łukasz Bondaruk, Jakub Kubiak, Adam Wiącek, Łukasz
  Degórski, Marek Kubis, and Paweł Skórzewski.
\newblock Spoken language corpora augmentation with domain-specific
  voice-cloned speech, 2024.

\bibitem{pelcra-tools}
Pelcra tools.
\newblock http://docs.pelcra.pl/.

\bibitem{mailabs}
The m-ailabs speech dataset.
\newblock https://github.com/imdatceleste/m-ailabs-dataset.

\bibitem{commonvoice}
R.~Ardila, M.~Branson, K.~Davis, M.~Henretty, M.~Kohler, J.~Meyer, R.~Morais,
  L.~Saunders, F.~M. Tyers, and G.~Weber.
\newblock Common voice: {A} massively-multilingual speech corpus.
\newblock In {\em Proc. 12th Conference on Language Resources and Evaluation},
  pages 4211--4215, 2020.

\bibitem{conformer}
Anmol Gulati, James Qin, Chung-Cheng Chiu, Niki Parmar, Yu~Zhang, Jiahui Yu,
  Wei Han, Shibo Wang, Zhengdong Zhang, Yonghui Wu, and Ruoming Pang.
\newblock Conformer: Convolution-augmented transformer for speech recognition.
\newblock In {\em Proc. Interspeech 2020}, pages 5036--5040, 2020.

\bibitem{transformer}
Ashish Vaswani, Noam Shazeer, Niki Parmar, Jakob Uszkoreit, Llion Jones,
  Aidan~N. Gomez, \L{}ukasz Kaiser, and Illia Polosukhin.
\newblock Attention is all you need.
\newblock In {\em Proceedings of the 31st International Conference on Neural
  Information Processing Systems}, NIPS'17, page 6000–6010, Red Hook, NY,
  USA, 2017. Curran Associates Inc.

\bibitem{flowmatching}
Yaron Lipman, Ricky T.~Q. Chen, Heli Ben-Hamu, Maximilian Nickel, and Matt Le.
\newblock Flow matching for generative modeling, 2023.

\bibitem{u-net}
Olaf Ronneberger, Philipp Fischer, and Thomas Brox.
\newblock U-net: Convolutional networks for biomedical image segmentation.
\newblock In Nassir Navab, Joachim Hornegger, William~M. Wells, and
  Alejandro~F. Frangi, editors, {\em Medical Image Computing and
  Computer-Assisted Intervention -- MICCAI 2015}, pages 234--241, Cham, 2015.
  Springer International Publishing.

\bibitem{rotary}
Jianlin Su, Murtadha Ahmed, Yu~Lu, Shengfeng Pan, Wen Bo, and Yunfeng Liu.
\newblock Roformer: Enhanced transformer with rotary position embedding.
\newblock {\em Neurocomputing}, 568:127063, 2024.

\bibitem{audiobox}
Apoorv Vyas, Bowen Shi, Matthew Le, Andros Tjandra, Yi-Chiao Wu, Baishan Guo,
  Jiemin Zhang, Xinyue Zhang, Robert Adkins, William Ngan, Jeff Wang, Ivan
  Cruz, Bapi Akula, Akinniyi Akinyemi, Brian Ellis, Rashel Moritz, Yael
  Yungster, Alice Rakotoarison, Liang Tan, Chris Summers, Carleigh Wood, Joshua
  Lane, Mary Williamson, and Wei-Ning Hsu.
\newblock Audiobox: Unified audio generation with natural language prompts,
  2023.

\bibitem{best-rq}
Chung-Cheng Chiu, James Qin, Yu~Zhang, Jiahui Yu, and Yonghui Wu.
\newblock Self-supervised learning with random-projection quantizer for speech
  recognition, 2022.

\bibitem{hifigan}
Jungil Kong, Jaehyeon Kim, and Jaekyoung Bae.
\newblock Hifi-{GAN}: generative adversarial networks for efficient and high
  fidelity speech synthesis.
\newblock In {\em Proceedings of the 34th International Conference on Neural
  Information Processing Systems}, NIPS'20, Red Hook, NY, USA, 2020. Curran
  Associates Inc.

\bibitem{specaug}
Daniel~S. Park, William Chan, Yu~Zhang, Chung-Cheng Chiu, Barret Zoph, Ekin~D.
  Cubuk, and Quoc~V. Le.
\newblock Specaugment: A simple data augmentation method for automatic speech
  recognition.
\newblock In {\em Proc. Interspeech 2019}, pages 2613--2617, 2019.

\bibitem{comparison-asr-augmentation}
Mina Huh, Ruchira Ray, and Corey Karnei.
\newblock A comparison of speech data augmentation methods using s3prl toolkit,
  2024.

\end{thebibliography}

\end{document}